# DNA BASE DETECTION USING TWO-DIMENSIONAL MATERIALS BEYOND GRAPHENE

BENJAMIN O. TAYO

*Department of Engineering and Physics*

*University of Central Oklahoma*

*100 N. University Dr.*

*Edmond, OK 73034*

*USA*
*btayo@uco.edu*

Graphene's success for nanopore DNA sequencing has shown that it is possible to explore other potential single and few-atom thick layers of 2D materials beyond graphene, and also that these materials can exhibit fascinating and technologically useful properties for DNA base detection that are superior to those of graphene. In this article, we review the state-of-the art of DNA base detection using 2D materials beyond graphene. Initially, we present an overview of nanopore-based DNA sequencing methods using biological and solid-state nanopores, and discuss several challenges that limit their use for single-base resolution. Then we outline the progress, challenges, and opportunities using graphene. Additionally, we discuss several potential 2D materials beyond graphene such as hexagonal boron nitride, elemental 2D materials beyond graphene, and 2D transition metal dichalcogenides. Finally, we highlight the potential of using van der Waals materials for advanced DNA base detection technologies.

## 1. Introduction

Biological materials such as deoxyribonucleic acid (DNA) and protein are polymers that are made from molecular building blocks called monomers. The structure and function of a biopolymer depend on the sequence of its repeating units. Any damage or alteration to one of these monomers could trigger mutations with long-term deleterious effects. For instance, DNA base damage, if not properly repaired, could trigger potentially carcinogenic changes. The ability to detect and discriminate DNA bases using simple and cost-effective methods is an important problem whose solution could lead to widespread deployment of DNA sequencing (e.g., in clinical applications and eventually in personalized medicine).

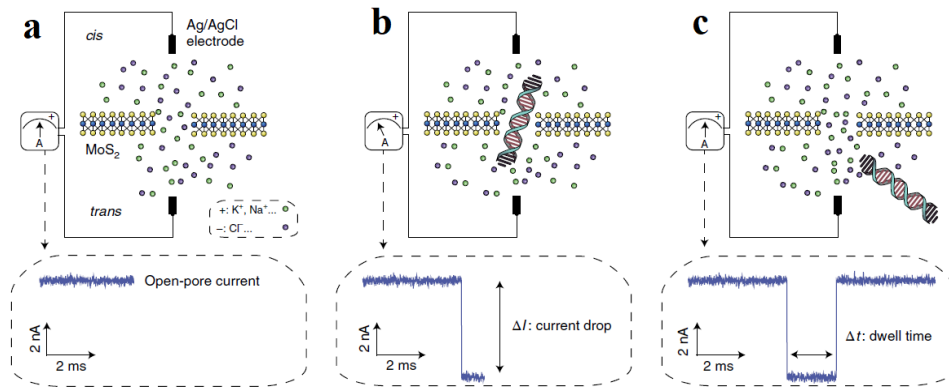

**Fig. 1. Nanopore-sensing principle. a**, Open-pore current caused by ions migrating in the electric field created by the Ag/AgCl electrodes. **b**, The negatively charged double-stranded (ds)DNA gets pushed through the nanopore by the high electric field at this location. Consequently, the current drops by a magnitude of ΔI, proportional to the size of the molecule. **c**, The dsDNA molecule exits the pore, and the current level returns to the open-pore value. The dwell time extracted from this translocation event is proportional to the length of the molecules. MoS$_2$ thickness and DNA molecules are not to scale. Adapted by permission from: Springer Nature Ref. 7, Copyright (2019).

Over the past two decades, nanopore-based DNA sequencing methods have been at the center of intense research as they continue to open up new avenues for fast and high-resolution DNA sequencing.[1-4]. In a typical nanopore sequencing experiment (see Fig. 1), individual single-stranded DNA (ssDNA) or double-stranded DNA (dsDNA) molecules are electrophoretically driven through the pore along with other molecules such as water and ions by an applied external electric field.[5,6] DNA translocation through nanopores would cause a sudden drop in the ionic current due to DNA occupation of the pore.[6,8] Because of the difference in the size of DNA bases and their interaction with the pore, the ionic current changes when each base is driven through. Two parameters extracted from the ion-blockade method that uniquely defines a base are the current drop and the drop duration time (dwell time), as shown in Fig. 2. The current drop is proportional to the size of the molecule and the dwell time is proportional to the length of the molecules. The modulated measured current, therefore, represents a direct reading of the DNA sequence. The translocation typically occurs at speeds of about $10^7$ bases/s.[8,9]

From an experimental point of view, nanopores can be fabricated using multiple of methods, such as electron-beam lithography (EBL) with reactive-ion etching (RIE), ion beam milling[4], dielectric breakdown[10], drilling with transmission electron microscope (TEM)[11] or laser beam fabrication.[12,13] An overview of the fabrication process is shown in Fig. 3.

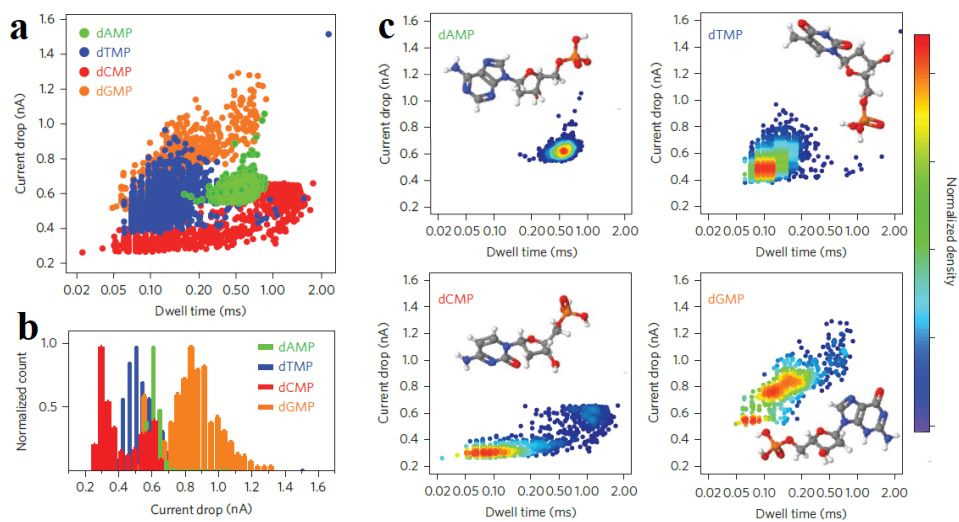

**Fig. 2. Identification of single nucleotides in a MoS$_2$ nanopore. a**, Scatter plots of nucleotide translocation events, showing distinguished current drops and dwell times for dAMP, dCMP, dTMP and dGMP. **b**, Normalized histogram of current drops for dAMP, dTMP, dCMP and dGMP. **c**, Density plots of single nucleotides in the MoS$_2$ nanopore. Adapted by permission from: Springer Nature Ref. 9, Copyright (2015).

Biological nanopores such as α-hemolysin[1] or other solid-state nanopores fabricated on membranes such as Si$_3$N$_4$,[4] SiO$_2$,[14] Al$_2$O$_3$,[15] and plastic[16] have been extensively used for DNA sequencing. For example, nanopores in silicon nitride membranes have been used to distinguish single- and double-stranded DNA,[17] and different polynucleotides;[18] and were able to detect DNA folding.[19] Furthermore, solid-state nanopores have also been shown to detect proteins,[20] and have been used to elucidate the dynamics of protein folding,[21-23] as well as extract the protein's shape in real-time.[24] While these methods have enjoyed tremendous success in the field of single-molecule analysis, they suffer from several challenges that limit their use for single-base detection:

(i) biological nanopores are very sensitive to temperature, PH, and applied voltage, making them unsuitable for practical applications,[3]

(ii) most biological nanopores such as α-hemolysin or other solid-state nanopores fabricated on membranes such as Si$_3$N$_4$ are typically more than 10 – 20 nm thick (which is equivalent to about 30 – 60 DNA bases),[25] which makes it difficult to detect individual bases-specific modulation in ion currents or transverse tunneling currents as multiple base pairs interact with the pore/gap simultaneously,[25,26]

(iii) fast speed of DNA translocation (10$^7$ bases/s in solid-state nanopores) makes single-base resolution using ionic currents challenging,[8] and

(iv) low bandwidth in recording of the ionic current.[27]

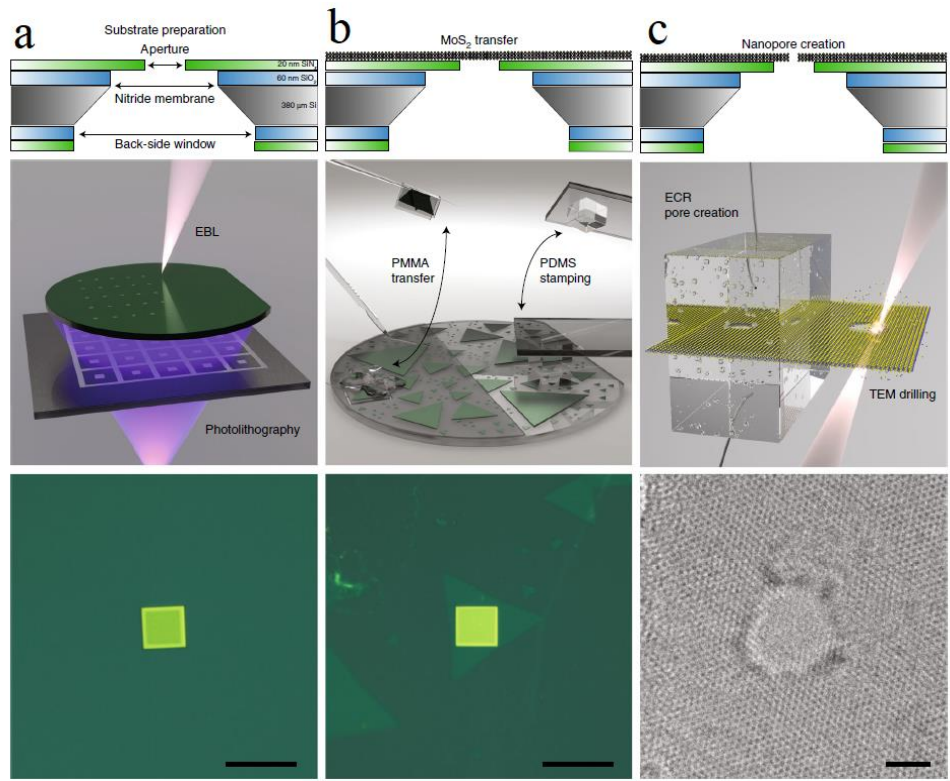

**Fig. 3. Overview of the nanopore fabrication process. a**, Substrate preparation. Top: a schematic of the finished device. Middle: illustration of the involved methods: EBL is used to pattern the aperture that is used to suspend the $MoS_2$ layer, whereas photolithography is used to create the backside opening for KOH etching. Bottom: an optical micrograph of the resulting freestanding silicon nitride membrane. Scale bar, 20 μm. **b**, Transfer of $MoS_2$. Top: a schematic of the device after $MoS_2$ has been transferred. Middle: an illustration of the two options for transferring the material: PMMA- and PDMS-assisted liftoff and alignment to the target substrate. Bottom: an optical micrograph of a single crystal of $MoS_2$ (triangle) transferred to the silicon nitride membrane. Scale bar, 20 μm. **c**, Drilling of a nanopore. Top: a schematic of the finished device. Middle: illustration of the two nanopore creation methods available to users: TEM drilling and ECR pore creation. Bottom: a TEM image of a drilled nanopore. Scale bar, 2 nm. ECR, electrochemical reaction; PDMS, polydimethylsiloxane; PMMA, poly(methyl methacrylate). Adapted by permission from: Springer Nature Ref. 7, Copyright (2019).

The emergence of ultrathin two-dimensional (2D) crystals such as graphene and transition metal dichalcogenides (TMDs) over the past two decades[28-30] has created new opportunities and potentials in the field of nanopore DNA sequencing. The single-layer nature of 2D materials is comparable to the size of the DNA base.[28,30] Hence these materials show strong promise to provide the necessary resolution at the single-base level.[31,32]

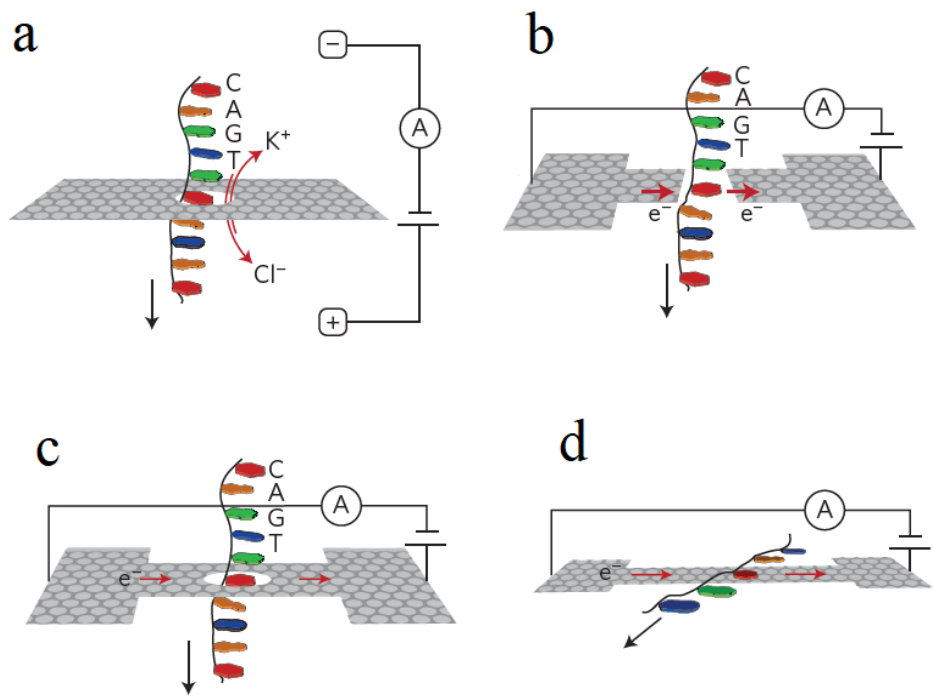

**Fig. 4. Four new concepts using graphene nanostructures for DNA sequencing. a**, Detection of changes in the ionic current through a nanopore in a graphene membrane due to the passage of a DNA molecule. **b**, Modulations of a tunnelling current through a nanogap between two graphene electrodes due to presence of a DNA molecule. **c**, Variations in the in-plane current through a graphene nanoribbon due to traversal of a DNA molecule. **d**, Changes in a graphene current due to the physisorption of DNA bases onto the graphene. Adapted by permission from: Springer Nature Ref. 38, Copyright (2016).

Graphene nanopores and nanogaps have been successfully used for DNA sequencing.[32,33] While the single-layer nature of graphene provides the optimal thickness (0.34 nm) for single-base resolution,[28] the major hindrance is the hydrophobic nature of graphene's surface. Because of the strong π-π interactions between graphene and the DNA, bases stick its surface[34,35] leading to a significant reduction in translocation speed due to pore clogging.[36] Furthermore, the coexistence of different bases on the surface and pore makes single-base discrimination difficult.[36] Another issue is the problem of orientational fluctuations of nucleobases during DNA translocation through a graphene nanopore. This can give rise to overlapping current contributions from different bases.[31] It has been shown that the ionic blockade signal shows noise for DNA translocation through a single-layer graphene nanopore.[35] The origin of this noise has been attributed to the atomic thickness of the pore. It is notable that a nanopore in a three-layer graphite structure, which has a thickness ~1 nm, shows a better signal-to-noise ratio (SNR).[35,37] The lack of a band gap in pristine graphene makes it undesirable for use in electronic-based detection modalities such as tunneling current or field-effect transistors (FETs).[8]

With all the disadvantages of graphene nanopores, most efforts have recently focused on 2D materials beyond graphene. Despite some challenges, graphene's success for nanopore DNA sequencing has shown that it is possible to explore other potential single and few-atom thick layers of 2D materials beyond graphene, and also that these materials can exhibit fascinating and technologically useful properties for DNA base detection that are superior to those of graphene. In the next sections, we discuss several potential 2D materials beyond graphene such as hexagonal boron nitride, elemental 2D materials beyond graphene, and 2D transition metal dichalcogenides (TMDs). We shall discuss about novel sensing approaches (see Fig. 4) such as electronic and optical modalities that have the capability to produce stronger signals compared to the conventional ion-current method.[38] Finally, we highlight the potential of using van der Waals (vdW) materials for advanced DNA base detection technologies.

## 2. Hexagonal Boron Nitride

Nanopores fabricated from graphene sheets can be made extremely thin and structurally robust to allow the identification of single nucleotides.[31–35] However, the graphene-based nanopore suffers from significant signal noises. The reason is that graphene is highly hydrophobic and sticks to the DNA strand, therefore trapping DNA bases during its translocation. This trap contributes to a high noise level for sequencing, making single base resolution quite challenging.[33,35,36] Recently, much attention has been given to hexagonal boron nitride (hBN) as an alternative to graphene for single base sequencing.[39-41] This is because hBN sheet exhibits the same honeycomb lattice structure as graphene (see Fig. 5), it has outstanding mechanical,[42] thermodynamic,[43] and electronic properties,[44] and it is easy to fabricate.[29,30] Despite its structural similarity to graphene, the electronic properties of hBN are drastically different from those of graphene: graphene is a gapless semimetal with the nonpolar nature of the homonuclear C−C bond, while hBN sheet is an insulator with the polar nature underlying the charge transfer between its constituent B and N atoms. hBN is less hydrophobic than graphene which can minimize the hydrophobic interaction that impedes the DNA translocation through the constructed nanopores. Moreover, the thickness of hBN is comparable to the spacing between nucleotides in ssDNA (0.32 – 0.52 nm).[45] It also shows other advantages over graphene in terms of its insulating property in high-ionic-strength solution and fewer defects made during the manufacturing process.[40] Experimental and theoretical studies show that hBN can reach spatial resolutions for DNA sequencing close to those of graphene. The fundamental properties of DNA nucleobases and hBN sheets remain unchanged upon adsorption, which suggests its promising application for DNA research.[46]

The above consideration shows that hBN is a nice alternative to graphene and could be used in combination with graphene for advanced functional devices for DNA sequencing with higher SNR.[41]

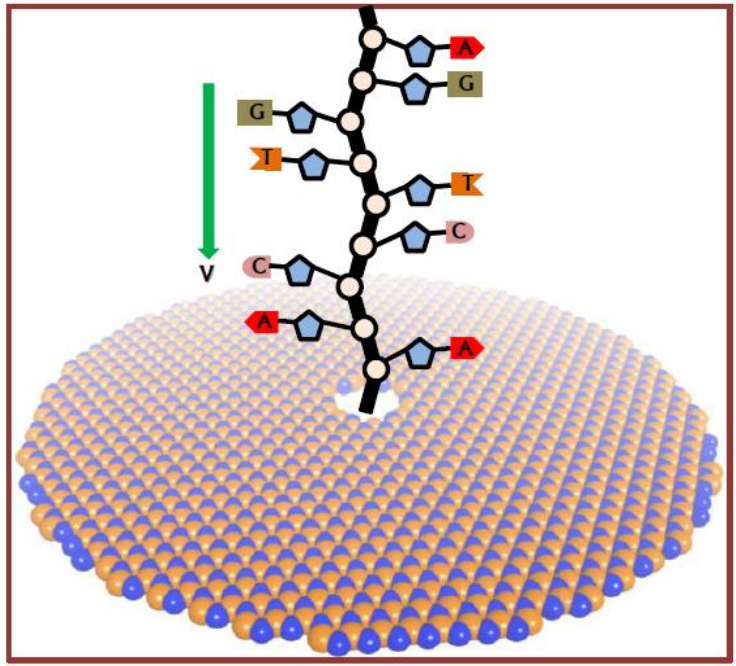

**Fig. 5.** Scheme of single-strand DNA (ssDNA) sequencing with hexagonal boron nitride (hBN) nanopore. The ssDNA is placed right above the hBN nanopore and perpendicular to the hBN sheet. The red, orange, pink and grey areas represent the adenine (A), thymine (T), cytosine (C) and guanine (G) nucleotides separately. Adapted from Ref. 39.

### 3. Elemental 2D Materials

Graphene remains the most studied elemental 2D material for DNA sequencing.[7,38] While tremendous success has been reported with the use of graphene for DNA sequencing, the major challenge is the stickiness of bases to graphene's surface. Other elemental 2D materials might provide equal or better properties than graphene. Hence more explorative work using other elemental 2D nanosheets as alternative to graphene is needed. Our group is currently performing computational studies to evaluate other elemental 2D materials such as silicene,[29,47] germanene,[29,48] and phosphorene (see Fig. 6).[28,49] Phosphorene, in particular, has a direct band gap and is hydrophilic.[28,49] Such properties are expected to resolve the problem of stickiness observed in graphene. Also, it can produce strong couplings with DNA bases, leading to larger SNR. Our studies will include all effects that can alter SNR such as pore/gap size, the geometry of pore/gap, edge geometry of the device (armchair or zigzag), and edge passivation. For instance, edge passivation of electrodes in graphene nanogap devices has been shown to improve coupling between electrodes and bases,[50,51] and slowing down the translocation speed of the DNA, allowing more time for the individual base to interact with the electrodes.[50] Also, employing different passivation mechanisms[52-54] such as hydrogenation, oxygenation, and hydroxylation will provide additional opportunities for tuning the signal during DNA sequencing.

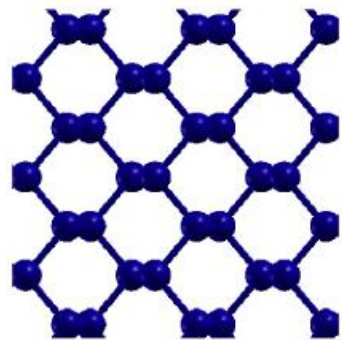

**Fig. 6.** Crystal structure of phosphorene (Black Phosphorus). From Ref. 28. Adapted with permission from AAAS.

## 4. 2D Transition Metal Dichalcogenides

These materials have desirable properties such as tunable direct energy band gap, and excellent mobility.[29,30] Moreover, they are easy to fabricate.[7,29] Because of their hydrophilic nature, they are anticipated to produce strong couplings with bases on the DNA. Such a strong coupling will enhance the tunneling currents leading to improved SNR.

Among the large family of TMDs, $MoS_2$ is the most widely studied for sequencing applications, mainly due to the ease of fabrication of $MoS_2$ devices.[7,29,55] Several theoretical and experimental studies have demonstrated sequencing using single-layer nanopores of $MoS_2$.[7-9,56,57] These studies reveal that $MoS_2$ performs better than graphene. For instance, improved SNR, non-stickiness of DNA to $MoS_2$ surface, and the presence of an intrinsic band gap makes it suitable for use in advanced sequencing devices such as FETs or the nanochannel device (see Fig. 7).[7,8] The figures below show electronic-base detection principles using nanosheets from $MoS_2$. The figures show that each base display a characteristic response due to its interaction with nanopore or the Armchair Nanoribbon (AMNR), showing the potential of advanced and efficient detection principles based on probing electronic current. The use of nanopores from 2D membranes from TMDs provides additional opportunities for manipulating pore chemistry. In $MoS_2$ nanopores, for example, the atoms in the nanopore exposed to DNA bases can be engineered with three configurations: Mo-terminated, S-terminated, and both Mo- and S-terminated. Studies have shown that Mo-terminated pores couple more strongly with DNA bases compared to the S-terminated pores.[7] Using an alloy such as $MoWS_2$ provides even more configurations: Mo-terminated, W-terminated, S-terminated, Mo- and S-terminated, W- and S-terminated, Mo- and W-terminated or Mo-, W-, and S-terminated. Thus, providing more opportunities for tuning the signal from the sensing device.

Despite all the success recorded with the use of $MOS_2$ for DNA sequencing, there are still fundamental limitations related to translocation speed that must be resolved.[25,26,58] $WS_2$ has also been investigated for sensing properties and it showed properties similar to $MoS_2$.[59] Moreover, more explorative work using other TMD nanosheets such as $MoSe_2$, $WSe_2$, $MoTe_2$, $WTe_2$, and $ReSe_2$ is essential, as this might help identify potential sensing materials that might enable ultrafast DNA sequencing. Our group is conducting research to investigate several 2D TMDs for use in high-resolution DNA sequencing devices.

Potentially, our research could be extended to include other 2D materials such as transition metal oxides.[30]

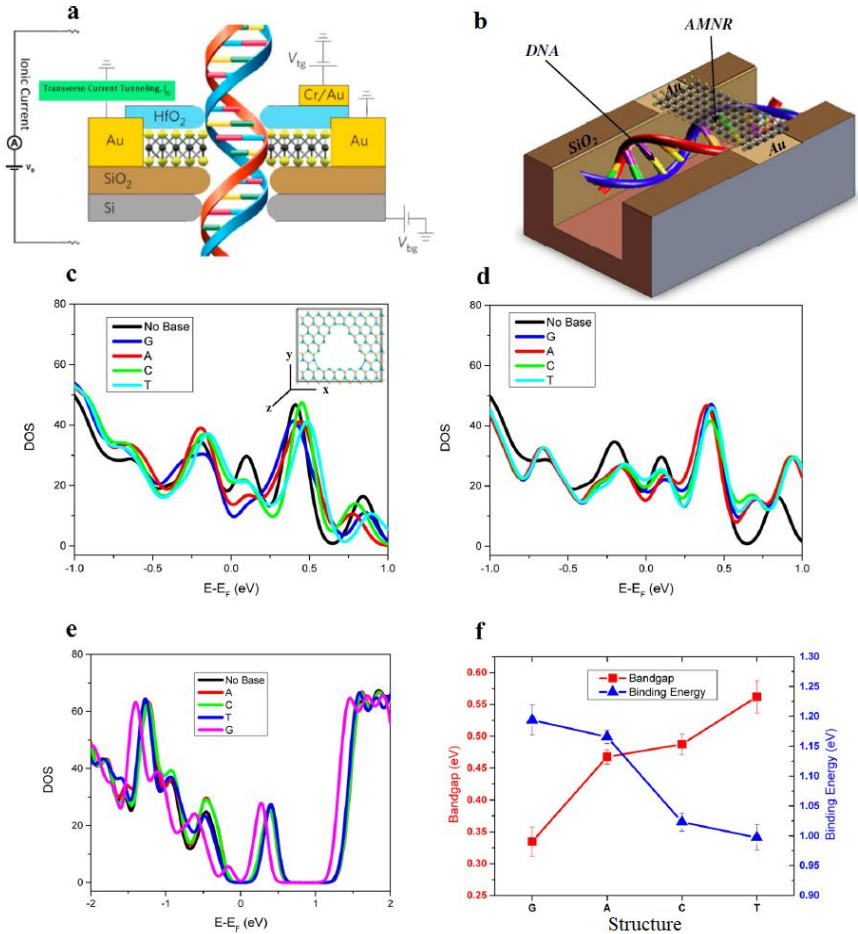

**Fig. 7.** a) Schematic view of a FET DNA sensing device. (b) Schematic device setup for Armchair Nanoribbon (AMNR) based DNA sensor. (c) horizontally (inset: molecular snapshot of Mo-only edge) and (d) at 45° angle in porous (Mo-terminated) $MoS_2$. (e) Total DOS and (f) band gap (left axis) and binding energy (right axis) of a pristine single-layer AMNR with each DNA base placed on the top of AMNR. Adapted with permission from Ref. 8. Copyright (2014). American Chemical Society.

## 5. Van der Waal Systems

In this section, we will focus on 2D-based vdW systems[28-30] and their potential for applications in DNA sequencing. vdW heterostructures formed by combining two or more single-layer 2D materials allow for a greater number of potential sensing materials (see

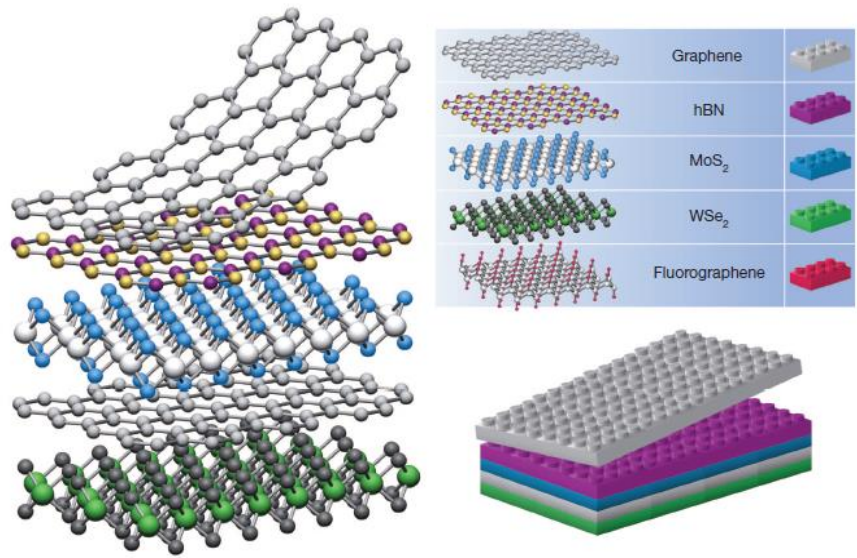

**Fig. 8. Building van der Waals heterostructures.** If one considers 2D crystals to be analogous to Lego blocks (right panel), the construction of a huge variety of layered structures becomes possible. Conceptually, this atomic-scale Lego resembles molecular beam epitaxy but employs different 'construction' rules and a distinct set of materials. Adapted by permission from: Springer Nature Ref. 30, Copyright (2013).

Fig. 8). Also, the synergetic effects produced when different materials are combined could lead to advanced detection properties that are superior to those of the individual materials. It has been shown that the ionic current blockade signal is prone to noise for DNA translocation through a single layer graphene nanopore (0.34 nm thickness).[35] Increasing the number of layers could enhance the SNR. For example, a nanopore in a three-layer graphite structure, which has a thickness of ~ 1 nm shows better SNR.[35,37] This observation indicates that the key to DNA base resolution could be found by using hybrid systems that consist of the stacking of single-layer 2D materials into vdW heterostructures. The graphene-$WS_2$ system, for example, has a thickness of about 1.4 nm, which makes it a superior material compared to graphene (thickness of 0.34 nm) and $WS_2$ (thickness of 1 nm) in terms of signal/noise intensity.[35,37] vdW heterostructures often have emergent properties that are superior to those of the individual building blocks.[28-30] When different 2D monolayers are stacked together to form a vdW heterostructure, the synergistic effects give rise to new emergent properties. For instance, interlayer coupling and ultrafast charge transfer across vdW interfaces have been demonstrated.[60] These emergent optical properties could even be used for DNA base detection principles based on modulations of optical properties in vdW systems due to their interaction with DNA bases. More explorative research studies for DNA sequencing using a variety of bi- and few-layer vdW

systems is essential. It would be interesting to compare the signals from vdW systems with the signals produced using the individual 2D building block materials, to see if there are synergistic advantages that could be exploited for DNA sequencing.

## 6. Conclusion and Perspectives

In summary, we have reviewed the current state-of-the art for DNA sequencing using 2D materials. The field of nanopore sequencing using 2D materials has already witnessed tremendous success. While single-base detection has been realized using these materials, most of the studies performed so far were based on probing ionic current variations caused by the blocking of nanopore by different bases during DNA translocation. The ionic current method is capable of achieving single-base resolution only at low translocation speeds due to low bandwidth for recording ionic current.[7,27] Also, to the best of our knowledge, among the large family of 2D materials, only graphene, hBN, $MoS_2$, and $WS_2$ have been investigated for DNA sequencing.[7] Even though $MoS_2$, $WS_2$, and hBN perform slightly better than graphene for DNA sequencing in terms of signal-to-noise ratio and non-stickiness of nanosheet to bases, there are still fundamental limitations related to the high translocation speed during DNA sequencing that has to be resolved.

At this stage, it is not clear what solid-state nanopore material will be able to meet the challenges for single-base resolution. Thus, it is critical to continue to carry out explorative studies to identify new nanopore materials that could potentially emerge as the best candidate material. To accomplish the objective of single-base resolution at high translocation speeds that might enable ultrafast DNA sequencing, much research is therefore still needed to explore the wide variety of potential 2D materials[61-63] including vdW heterostrcutures. Novel detection principles that are electronic-based such as the tunneling current or FET modalities must be investigated as well. These studies might reveal potential materials that are superior to graphene and $MoS_2$, and that are capable of producing single-base resolution at practical translocation speeds.

**Acknowledgments**

The author wishes to thank the College of Mathematics and Science at the University of Central Oklahoma for the CURE-STEM research funds, and the office of Research and Sponsored Programs for the 2019 Faculty Research Incentive Award. The author also acknowledges useful discussions with Dr. Chinedu E. Ekuma (Lehigh University).